\documentclass[conference]{IEEEtran}
\IEEEoverridecommandlockouts
\usepackage{cite}
\usepackage{amsmath,amssymb,amsfonts}
\usepackage{algorithmic}
\usepackage{graphicx}
\usepackage{textcomp}
\usepackage{xcolor}
\usepackage{multirow}
\usepackage{subfigure}

\def\BibTeX{{\rm B\kern-.05em{\sc i\kern-.025em b}\kern-.08em
    T\kern-.1667em\lower.7ex\hbox{E}\kern-.125emX}}
\begin{document}

\title{An investigation of modularity for noise robustness in conformer-based ASR

\thanks{Work done whilst the first author was at Telepathy Labs.
This work was partly funded by NCCR Evolving Language, a National Center of Competence in Research, funded by the Swiss National Science Foundation grant number 180888.}
}

\author{\IEEEauthorblockN{Louise Coppieters de Gibson}
\IEEEauthorblockA{\textit{Ecole Polytechnique Fédérale de Lausanne}
Lausanne, Switzerland \\
\textit{Idiap Research institute}
Martigny, Switzerland \\
louise.coppieters@idiap.ch}

\and
\IEEEauthorblockN{Philip N. Garner}
\IEEEauthorblockA{\textit{Idiap Research institute}\\
Martigny, Switzerland \\
pgarner@idiap.ch}

\and
\IEEEauthorblockN{Pierre-Edouard Honnet}
\IEEEauthorblockA{\textit{Telepathy Labs} \\
Zurich, Switzerland \\
pe.honnet@telepathy.ai}
}  

\maketitle
%

%
\begin{abstract}
Whilst state of the art automatic speech recognition (ASR) can perform well, it still degrades when exposed to acoustic environments that differ from those used when training the model. Unfamiliar environments for a given model may well be known a-priori, but yield comparatively small amounts of adaptation data.  In this experimental study, we investigate to what extent recent formalisations of modularity can aid adaptation of ASR to new acoustic environments.  Using a conformer based model and fixed routing, we confirm that environment awareness can indeed lead to improved performance in known environments.  However, at least on the (CHIME) datasets in the study, it is difficult for a classifier module to distinguish different noisy environments, a simpler distinction between noisy and clean speech being the optimal configuration.
The results have clear implications for deploying large models in particular environments with or without a-priori knowledge of the environmental noise. 
\end{abstract}
\section{Introduction}
With the advent of ever larger transformer-based architectures, the necessary compute power to train and infer with state of the art speech recognition models keeps increasing. Databases also grow in size and cover more modalities due to an increasing interest for multimodal and multitask modeling.
This leads to an exponential increase of the number of model parameters~\cite{sevilla2022compute} as well as amount of data required to train these models~\cite{epoch2022trendsintrainingdatasetsizes}. 

These larger models not only require large computational resources to be trained, they can also suffer from negative interference and catastrophic forgetting~\cite{ramasesh2021effect}.
Inspired by biological systems~\cite{sporns2016modular}, modularity has been applied in machine learning models for decades~\cite{jacobs1991adaptive, jordan1994hierarchical}. Recently, the concept of modularity has become popular again, especially for large models. Modularity can be introduced at different levels of a model and with different approaches, depending on the task and model structure~\cite{pfeiffer2023modular}. A modular structure allows to train only task-selected experts of a  model instead of the whole model. Irrelevant experts for a given task are thus not trained, which saves computational power and avoids catastrophic forgetting.
In the ASR field, mixture of experts (MoEs) have been shown to improve accuracy for different tasks by increasing the model size, while keeping the same computation power~\cite{you2021speechmoe, you2022speechmoe2, hu2023mixture}. 

This work focuses on introducing modularity in conformer-based ASR models to handle speech in different types of noise environment. By adding modularity at the conformer block level, we allow the model to learn different conditions and exploit this information to improve its performance on both noisy and clean speech. We hypothesize that using different experts for different noisy types of speech will enhance the ASR performance of each type of noisy speech. Introducing modularity in the beginning of the model would make the model more robust to different types of noisy speech with fixed routing. We then explore learned routing, where we observe similar performance but with a model better suited to handle real-life situations. 
We demonstrate the effectiveness of the approach on the task of speech recognition in noisy environments. Using different experts for clean and noisy speech outperforms the standard conformer, however adding more granularity inside the noisy data class by separating the different types of noise does not improve the performance further. Another finding is that our modular models tend to be trained faster than the baseline.

This paper first reminds the theory behind modularity and conformers in the Section 2. Then Section 3 introduces our proposed method to tackle challenging noise environments for ASR. In Section 4, we report our experiments and findings. We conclude in Section 5.

\section{Background}
\subsection{Modular networks}
A modular neural network has three specific components: functional blocks or \emph{experts}, a routing mechanism to select  the right experts, and an aggregator that combines the outputs of those experts. 

Functional blocks can be implemented in different ways:
\begin{itemize}
    \item It can be a composition of parameters such as sparse subnetworks, where a small number of parameters are pruned to be trained for each specific task~\cite{ansell2021composable} or low-rank modules such as LoRA~\cite{hu2021lora}. 
    \item It can be obtained through an input composition where the input is concatenated with specific parameters~\cite{li2021prefix}.
    \item It can be a function composition where a whole block is duplicated to act as different experts. \cite{rosenbaum2020dynamic} 
\end{itemize}
A routing mechanism is needed to select an expert. This routing mechanism can either be fixed if the expert selection is known from the data (for example in multitask learning~\cite{ruder2017overview}), or it can be learned when the routing information is not available. In case of learned routing, several challenges arise such as module collapse or training stability~\cite{rosenbaum2019routing}.

\subsection{Conformer}
The conformer architecture, introduced by Gulati et al.~\cite{gulati2020conformer} is a stack of conformer blocks. One such block is composed of two feedforward layers (one at the front and one at the end), one transformer layer and one convolutional layer. By design, it combines the advantages of both CNNs~\cite{sainath2013improvements} and  transformers~\cite{vaswani2017attention}: CNNs primarily capture local contextual information and dependencies, while self-attention captures more global context. 



\section{Method}



In this section we describe our two main contributions, namely the introduction of fixed routing and learned routing in the conformer architecture.

\subsection{Fixed routing}
Using a fixed routing mechanism implies knowing the condition in advance. When this is possible, a simple routing mechanism dictated by an input parameter can be set in place. This experimental setup gives us two keys: first it shows if using modularity to distinguish noisy and clean speech enhances the global performance. Second it can be used as pretrained model for a learned routing mechanism.  

\begin{figure*}
    \centering
     \subfigure[]{%
        \includegraphics[width=0.25\linewidth, trim = {0.8cm 0.8cm 0.8cm 0.8cm}]{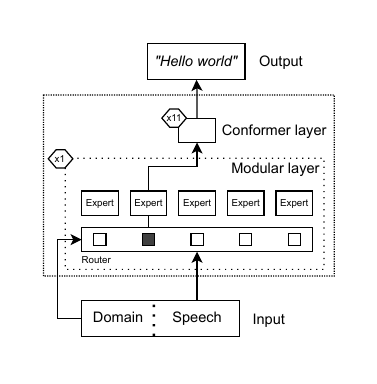}
        \label{fig:fixed_routing}}
    \hspace{1cm}
     \subfigure[]{
        \includegraphics[width=0.60\linewidth, trim = {1cm 0.8cm 1cm 0.8cm}]{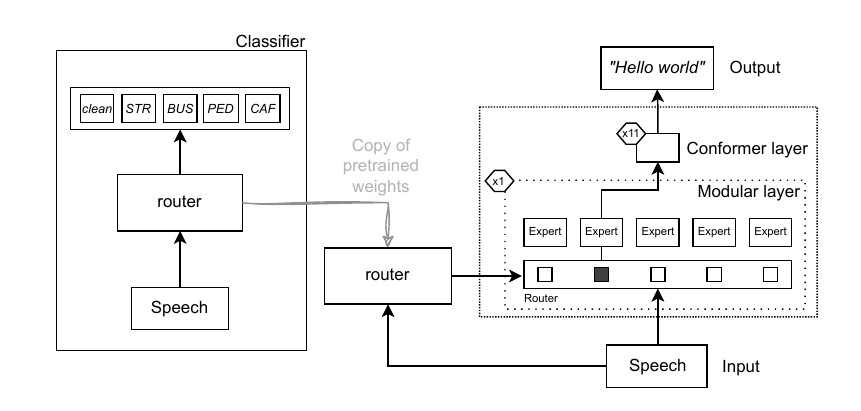}
        \label{fig:learned_routing}}
\caption{Different architectures used for the two different routing mechanisms: (a) Fixed routing architecture (b) Learned routing architecture}\label{fig:animals}
\end{figure*}

We propose to introduce modularity through experts at the conformer block level as illustrated in Figure~\ref{fig:fixed_routing}: every expert in a modular layer is a full conformer block. To keep the same amount of computations between the baseline and the modular approaches, the router chooses exactly one expert for each utterance. The aggregator then composes the batch in the right order after the modular layer.
 
\subsection{Learned routing}
The information about noise is not always available with the input. In daily situations, one can be exposed to outside noise or be in a noise free environment, but the model does not have access to the information. When this is the case, fixed routing cannot be used and the routing has to be inferred from the input signal. 

Expert modules can only start to differentiate when the router has learned a consistent pattern. This gives two main paths to train a learned router with an ASR model: 
train everything together from scratch, or first pretrain the router before integrating it with the ASR.

Training everything together from scratch implies setting up a constraint to diversify the choice of the router output in the beginning of the training. This forces the ASR experts to first learn a general solution, before starting to have a consistent routing.

The use of a pretrained router offers the advantage to be less computationally intensive, but it requires to create a parallel classification pipeline to pretrain the router. 
Pretraining the router avoids this forced diversification when training the ASR. Moreover, one can decide to freeze the router for some time while training the ASR and to unfreeze and train it jointly with the ASR in the next phase. In this paper, we used the second approach, illustrated in Figure~\ref{fig:learned_routing}.


\section{Experiments}

\subsection{Dataset}
Our experiments are carried out on the CHiME4 dataset~\cite{vincent20164th}, which is based on the wall street journal dataset. The clean part of the data is a combination of WSJ0 and WSJ1. The noisy part of the dataset consists of two components: a simulated portion (artificially mixed clean utterances with noisy background) and a real portion (speech recorded in noisy environment). The noises used in this dataset for training, validation and testing are drawn from daily life environments such as buses, cafés, streets and pedestrian areas. The same utterances are recorded in those different conditions. The data distribution is summarised in Table~\ref{tab:chime4}. In comparison to the initial dataset distribution, we added clean data in the test set to evaluate the model performance on a clean subset besides the noisy environments.
\begin{table}[h]
\caption{CHiME4 dataset summary with the number of utterances per subset. The noisy dataset contain the four types of noisy environments: bus, café, street and pedestrian area.}
\centering
\begin{tabular}{lllll}
\hline
      & clean  & noisy simu & noisy real & total\\
\hline
train & 37,416 & 42,828     & 9,600       & 89,844\\
dev   & -      & 1,640      & 1,640       & 3.280\\
eval  & 1,206  & 1,320      & 1,320       & 3,846\\ 
\hline
\end{tabular}
\label{tab:chime4}
\end{table}

\subsection{Baseline and framework}
Our implementation is based on the WeNet framework, an open-source toolkit used for streaming and non-streaming end-to-end speech recognition~\cite{yao2021wenet, zhang2022wenet}. 
The baseline model is a 12-layer conformer encoder with a 6-layer transformer decoder. At every layer of the encoder, the attention module has 4 attention heads.  The results of the baseline experiment are summarised in Table~\ref{tab:baseline}\footnote{Note that the results reported by the authors on github differ from what we were able to reproduce, especially for the SE condition}.

\begin{table}[h]
\caption{Baseline results: the results are computed using two decoding methods: 'ctc' for ctc beam search and 'att' for attention rescoring. Five subsets are chosen: real dev (RD), simu dev (SD), real eval (RE) and simu eval (SE) according to table \ref{tab:chime4}. }
\centering
\begin{tabular}{llllll}
\hline
        & clean  & RD & SD & RE & SE\\
\hline
ctc   & 17.73 & 20.91     & 22.48      & 30.85 & 53.66\\
att.  & 16.44 & 19.76     & 21.63      & 29.69 & 52.98\\
\hline
\end{tabular}

\label{tab:baseline}
\end{table}

\subsection{Fixed Routing}
In the fixed routing experiment we explored the impact of modularity when using different numbers of experts and expert layers: 
The expert choices ares:
\begin{itemize}
    \item 2 experts: clean, noise
    \item 3 experts: clean, simulated noise, real noise
    \item 5 experts: clean, bus, café, pedestrian area, street
\end{itemize}
For two experts, we also vary the number of layers which become modular: we experimented with 1, 2 and 3 modular layers. In addition, we also test the network behaviour when introducing the modularity only on the second or third layer rather than on the first layer of the network.
The routing path is appended to the beginning of the input waveform to provide the router with the domain information.

There exists an interesting trade-off between the number of experts that can be trained on specific data and the amount of data that each expert sees during training. If the data is diversified over the different experts, the more experts, the better the model will be adapted to that specific type of data. On the other hand if different types of data are too similar to be differentiated, the more experts, the less data each expert will receive to adapt its weights.

The results show that using modularity at the conformer block level outperforms the baseline both on clean and noisy speech. For clean speech, we achieve between 36.3\% and 40.4\% relative WER reduction, while for noisy speech the improvements lie between 6.1\% and 15.1\% for all types of noise except the simulated evaluation test set, where the baseline results differ from the rest of the results.

This means that specialising one layer to differentiate noise from clean environment enables the model to handle the two data types within different experts, which improves the general ASR performance.
Going further into the implementation details, the results show that using 2 experts outperforms other settings most of the time on the CHiME4 dataset, which is probably linked to the trade-off  discussed earlier.


\begin{table}
\caption{Results of fixed routing with attention rescoring decoding method.}
\centering
\begin{tabular}{p{0.8cm}p{1.1cm}llllll}\hline
experts  & mod. layer & clean  & RD & SD & RE & SE\\

\hline
\multicolumn{2}{l}{ Baseline} & 16.44 & 19.76     & 21.63      & 29.69 & 52.98\\
 \hline
2 & 1    & 10.02 & \textbf{16.77} & 19.93 & 26.28 & 27.33\\
2 & 1 - 2 & 10.43 & 16.98 & 19.74 & 26.22 & 27.25\\
2 & 1 - 2 - 3 & 10.16 & 17.99 & 20.30 & 26.69 & 28.03\\
3 & 1 & 10.47 & 18.12 & 19.63 & 27.21 & 27.76\\
5  & 1 & 10.30 & 17.46 & 19.89 & 26.82 & 28.25\\
2 & 2 & 10.18 & 16.99 & \textbf{19.59} & \textbf{26.05} & \textbf{26.85} \\
2 & 3 & \textbf{9.81} & 16.99 & 19.63 & 26.27 & 26.88 \\
\hline
\end{tabular}
\label{tab:fixed_routing}
\end{table}

\subsection{Learned routing}
The learned routing mechanism is divided into two parts: the router classifier and the ASR model (see Figure~\ref{fig:learned_routing}).

\subsubsection{Router classifier}
We first train a classifier to predict the target classes, which will become our pretrained router in the next stage. We opt for a simple architecture which consists of 3 CNN blocks. The goal of the router classifier is to predict the noise type of an input waveform. Since in the ASR model the conformer receives the waveform after the feature extraction,  the router input can be the same.
The different classes are the same as for our fixed routing experiments. 
An example for 5 experts is represented on the left side of Figure~\ref{fig:learned_routing}.

\begin{figure}
    \centering
    \begin{tabular}[t]{cc}
    \subfigure[]{
    \centering
    \includegraphics[width=0.35\linewidth]{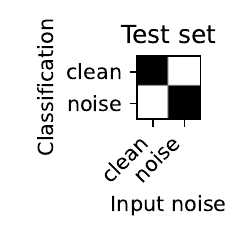}
    } 
        &
    \multirow{2}{*}{
    \subfigure[]{
    \centering
    \includegraphics[width=0.5\linewidth,trim = {1cm 0 0 3cm}]{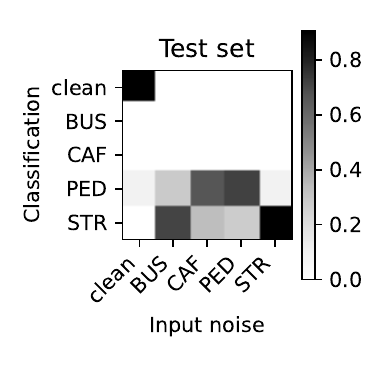}
    }}
     \\
    \subfigure[]{
    \centering
    \includegraphics[width=0.35\linewidth,trim = {0.5cm 0cm 0.5cm 0cm}]{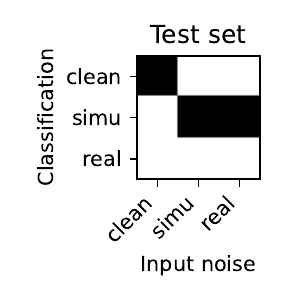}
    }


    \end{tabular}
    \label{fig:confusion_matrices}
    \caption{Confusion matrices for the different number of experts: (a) for 2 experts, (b)  5 experts and (c) 3 experts. The x-axis represent the type of domain we have at the input and the y-axis the output of the network.}
\end{figure}

The confusion matrices obtained for the different classifiers are shown in Figure~\ref{fig:confusion_matrices}.
For each figure, the true classes are given on the x-axis and the prediction is given on the y-axis of the matrix. 
For two experts (Figure~\ref{fig:confusion_matrices} (a)), the two classes are clearly distinct. For three experts (Figure~\ref{fig:confusion_matrices} (c)) the classifier is not able to make the distinction between real and simulated speech, but clean speech is clearly distinguished from noisy speech. 
Finally for five classes (Figure~\ref{fig:confusion_matrices} (b)), the classifier is able to distinguish some noises, but there is still some confusion between the different types of noise. This means that the classifier is not able to separate the 4 different classes based on speech features. Interestingly, it groups some noises together: on the one hand 'human activities' (café and pedestrian area)  and on the other hand 'car noises' (bus or street area) adds up as noise to the speech signal.

\subsubsection{ASR}
For the full speech recognition model with learned routing, the weights of the router classifier are loaded into the router part of the model (see Figure~\ref{fig:learned_routing}). We then did two different experiments: in the first one, we kept the weights of the router fixed during the whole ASR training, in the second one we kept the weights of the router fixed for the first five epochs and let the router then free to train (table \ref{tab:learned_routing}). 

The results are similar to the ones obtained in the fixed routing experiments. This is due to the use of number of experts corresponding to what this dataset is able to differentiate amongst the different experts after feature extraction. 

We then analysed what the router tended to learn when it is free to train. For 2 and 3 experts, after we unfreeze it, the router tends to transfer all the incoming data to the noise adapted expert, while for 5 experts the model keeps the different experts separated. The choice of the 'noise-robust' expert is probably due to the fact that this expert is trained to handle noisy speech and easily adapts to less noisy environments, while the other one only adapts to clean speech. 

\begin{figure}
    \centering
    \includegraphics[width=\linewidth]{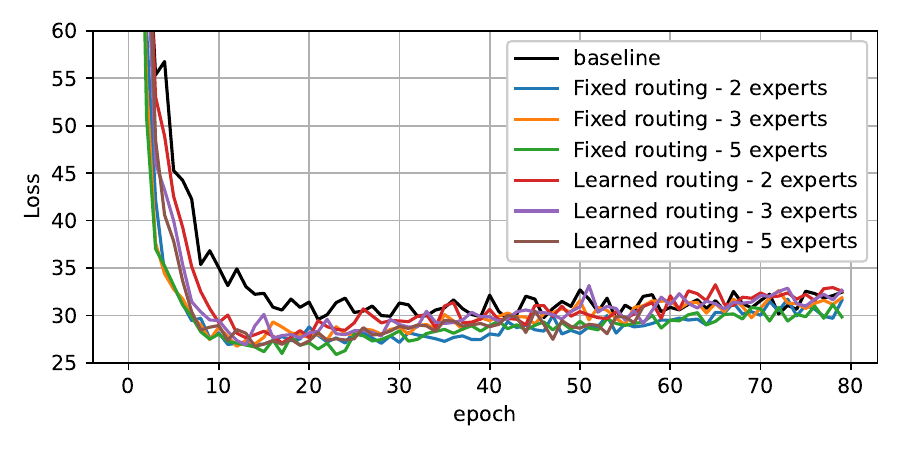}
    \caption{Loss function of different experiments: baseline and fixed and learned routing.}
    \label{fig:loss_learned_routing}
\end{figure}

Further analysing the loss function (see figure \ref{fig:loss_learned_routing}) shows that using routing helps the model to converge faster: after approximately 20 epochs, while the baseline experiment takes more time (about 25-35 epochs). This is reflected in the final model as we take the average of weights from the best 10 models, based on the validation loss. The final baseline model uses model checkpoints from epochs between 21 and 54, while for all the models where we introduce modularity, the final model is the average of checkpoints between the epochs 10 and 35\footnote{One exception was observed for the learned routing with 5 experts, with one outlier checkpoint being epoch 47.}. The implication is that we are able to reach better performance in a reduced training time and therefore less computing power.

\begin{table}
\caption{Results of learned routing mechanism}
\label{tab:learned_routing}
\centering
\begin{tabular}{p{0.8cm}p{1.1cm}llllll}\hline
experts  & number of fixed epochs & clean  & RD & SD & RE & SE\\
\hline
\multicolumn{2}{l}{Baseline} & 16.44 & 19.76     & 21.63      & 29.69 & 52.98\\
\hline
2             & 80            &   9.97 & 17.47 & 19.93 & 26.89 & 28.14 \\
3             & 80            & 9.90  & 16.96 & 19.46 & 26.12 & 26.78 \\
5             & 80            & 10.25 & 17.08 & 19.63 & 26.48 & 27.38\\
\hline
2            & 5            & 10.56 & 17.02 & 19.34 & 25.97 & 26.14\\
3             & 5            & 10.22  & 17.70 & 19.34 & 26.60 & 27.85 \\
5             & 5            & 9.91 & 17.17 & 19.66 & 26.48 & 26.96 \\
\hline
\end{tabular}
\end{table}
\section{Conclusion}
In this paper we examine the effectiveness of modularity on the CHiME4 dataset. We introduce modularity at the conformer block level and two routing options are explored: fixed and learned routing.

The fixed routing approach demonstrates that using modularity consistently outperforms the baseline across all conditions and configurations. However, dividing the data between two experts yields better results than using three or five. This suggests a trade-off: when input signals are distinctly different at the point where modularity takes place, using separate experts for each type improves overall performance. However, if the signals are similar, multiple experts may end up learning the same task, effectively reducing the data each expert processes.


We explored learned routing via a classifier-based router, which is pretrained before integration into the ASR system. This classifier shows that noisy speech is more challenging to differentiate after feature extraction, leading to a final division into two or three experts that distinguish between clean and noisy speech. This also points that noise distinction better works on SNR level than on the type of noise.

We also showed that the introduction of modularity allows for faster training, meaning reduced computational resources.

Future work may explore techniques to use fully learned routing without target classes. 
This approach can bring up other distinguishable elements helpful to ASR. 


\bibliographystyle{IEEEbib}
\bibliography{refs}

\end{document}